\documentclass[11pt,twoside]{article}  
\usepackage{adassconf}

\begin{document}   

%
%
%

\paperID{O4a-2}

%
%
%
%

\title{Knowledge Discovery Framework for the Virtual Observatory}

%
%
%

\author{Brian Thomas, Edward Shaya, Zenping Huang, Peter Teuben}
\affil{Department of Astronomy, University of Maryland, College Park, MD, 20742}

%
%

\contact{Brian Thomas}
\email{thomas@astro.umd.edu}

%
%
%
%
%

\paindex{ Thomas, B.}
\aindex{ Shaya, E. }     
\aindex{ Huang, Z. }     
\aindex{ Teuben, P. }     

%
%

\authormark{ Thomas et al.}

%
%



\begin{abstract}          
We describe a framework that allows a scientist-user to easily query for
information across all Virtual Observatory (VO) repositories and pull it back
for analysis. This framework hides the gory details of meta-data remediation
and data formatting from the user, allowing them to get on with search,
retrieval and analysis of VO data as if they were drawn from a single source
using a science based terminology rather than a data-centric one.
\end{abstract}

%
%

\section{The problem with the VO}
A key problem facing the Virtual Observatory (VO) is that the search for and fusion of VO data 
for scientific use still requires a human. The reason for this is easy to understand: the VO
includes many heterogeneously, and incompletely (from a semantic standpoint) described data. 
Heterogeneous description arises from the differing database schema in which the data are held.
UCD's (see [1]) have been used to help solve semantic description of the data, however they 
only label the semantic meaning of the columns within tables, but they do not 
label overall content of the table, nor is there any understanding of how one table relates to
another.

Thus the scientist-user of the VO must know the nature of the schema at a given
repository, tailor their query to match it and then fuse/transform the data themselves, using 
both their knowledge of the field of astronomy, and the data at the given repositories to achieve 
the resulting dataset they desire to do their research.

This is not the manner in which most (or all!) users of the VO would like to proceed to do 
research! Instead of thinking about the nuts and bolts of locating 
and downloading and combining the data they would prefer to ask 
a question of the VO like: ``find stars with measured IR magnitudes which have been observed inside a spiral galaxy arm'' and the machine will handle the dirty work of search and fusion. 

\section{The Solution: semantic interoperability}
One means to solve this problem is to 
introduce a knowledge discovery framework around the existing data repository structure. 
This framework will provide what is called ``semantic interoperability'',
which may be defined as 
\begin{quote}
A dynamic capability that allows a machine to infer, relate, interpret and classify 
the implicit meanings of digital content without human involvement.\footnote{Paraphrased
from the definition in [2]} 
\end{quote}
An astronomical example of semantic interoperability: the machine is able to 
determine which (previously unidentified) astrometry data are about ‘Pluto’ by 
referencing an ephemeris of Pluto's orbit. Another example: the machine uses 
extant properties of spiral galaxy data such as I-band magnitude, inclination 
and rotational velocity to calculate a new property of distance using the 
Tully-Fisher equation. 

\section{Building a knowledge discovery framework}

We have developed a design for a knowledge discovery framework for the VO.
The framework consists of a semantic layer which allows mapping of semantic information to
existing data, an ontological model of astronomical objects which includes 
details of the scientific relationships that exist between objects, and 
one or more tools to aid the human in utilizing the layer and ontology to
discover, retrieve, and transform VO data.

\begin{figure}
\epsscale{.40}
\plottwo{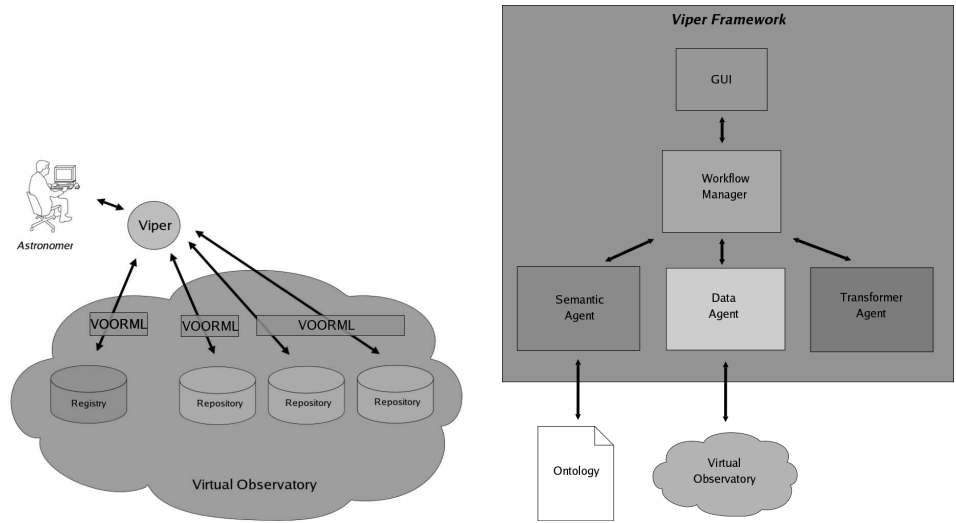}{O4a-2_2.epsi} 
\caption{Summary of knowledge discovery framework. {\it Left plot}: The relationship between Human, VOORML, VO repositories and Viper. {\it Right plot}: major components of Viper tool.} \label{O4a-2-fig-1}
\end{figure}

Figure~\ref{O4a-2-fig-1} provides an overview of this framework which comprises
a semantic layer (VOORML, VO Object-Relational mapping layer), and a tool, Viper, 
which intermediates between the semantic layer and the human.

Below we provide brief detail of parts of the framework.

\subsection{User tool : Viper}

\begin{figure}
\epsscale{.40}
\plotone{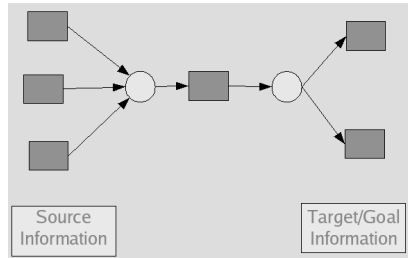}
\caption{Viper workflow representation. Squares represent semantic objects and circles 
indicate relationships between them. The workflow allows the user 
to design information flows wherein ``sources'' at repositories are retrieved, and then transformed 
as desired into ``goal'' form.}
\label{O4a-2-fig-4}
\end{figure}

This tool is designed to coordinate and aid the human in the tasks of discovery, 
retrieval and transformation of VO data. Viper consists of a number of major components 
including a semantic agent which coordinates inference made
against the ontology, a data agent which coordinates search and retrieval requests to
the semantic layer, and a transformer agent which acts on retrieved data to change it to
the desired semantic state (as well as perform data fusion). A workflow manager 
coordinates activities between the respective agents. 
 
Viper interacts with the human via a graphical workspace (similar to that shown in figure 
\ref{O4a-2-fig-4}) in which objects (squares) are interrelated with known relationships 
(circles). A menu (not shown) allows the user to drop and drag either of these types of 
semantic term into the workspace, and the tool prevents illegal semantic combinations 
(as determined by the ontology). Because objects/relationships are defined in the ontology,
Viper may be used to discover relationships the user is not aware of, and will then alter the 
workspace view appropriately to show them.

Viper, and the agents, are written in Java, and make use of Jena ([3]) and Pellet ([4]) 
to respectively represent, and reason on, the ontology. 

\subsection{Ontology}

Our ontology presently consists of greater than 1000 objects related to astronomy. We include
astronomical objects and phenomena as well as scientific relationships between them (such as
Kepler's Laws, the conversion between magnitude and an energy 
flux and so on). The ontology itself is serialized in OWL ([6]) and is 
realized as a number of sub-domains (each in a separate file) which include definitions of
physical measurements/quantities, geometry, instruments, units, physics and statistics.

\subsection{Semantic layer : VOORML}

\begin{figure}
\epsscale{.40}
\plotone{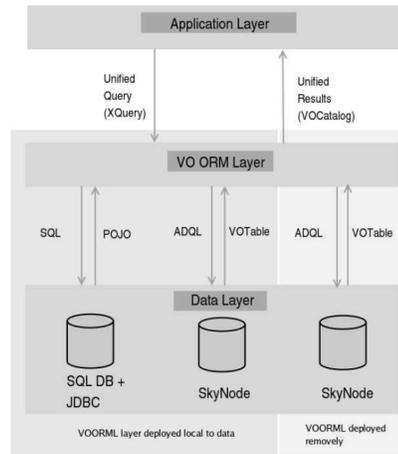}
\caption{{\it\b Detail of interaction between application, VOORML and data layers.}} \label{O4a-2-fig-3}
\end{figure}

This software provides a mapping between the data in the repository 
and semantic terms which are described by the ontology (figure~\ref{O4a-2-fig-3}).
In a nutshell, the layer uses a simple collection/quantity-based datamodel (see VOCatalog, [7]) 
to tag the semantic meaning of data at a repository. This model allows the layer to
understand how to map the parts of a query, framed in semantic terms, into a localized
query (in SQL). The model also directs the layer on how to reassemble matched data
back into the uniform, semantic model (again using VOCatalog) before it is returned.
Presently this software only will work upon a JDBC interface, but in the future we hope 
to make the simple translation to ADQL so that SkyNodes ([8]) may be described as well. 
Finally, there is no reason this software need be deployed locally at the site of the 
repository.

\section{Summary}

It is impossible to describe the entire system in detail here. See our website ([9]) for
further description and updates on progress of this work.

\end{document}